\documentclass[11pt,twoside]{article}

\usepackage{asp2004}
\usepackage{lscape}
\usepackage{amsmath,amssymb,amsbsy}
\usepackage{graphicx}

\begin{document}
\title{The Nature of Light Variations of the Helium Strong Chemically Peculiar
Star HD 37776}   
\author{Ji\v{r}\'{\i} Krti\v{c}ka$^1$, Zden\v ek Mikul\'a\v sek$^1$,
        Juraj Zverko$^2$, and Josef \v Zi\v z\v novsk\'y$^2$}
\affil{$^1$Institute of Theoretical Physics and Astrophysics,
Masaryk University, Kotl\'a\v{r}sk\'a 2, CZ-611\,37 Brno, Czech
Republic, krticka@physics.muni.cz}
\affil{$^2$Astronomical Institute of the Slovak Academy of Sciences,
SK-059\,60 Tatransk\'{a} Lomnica, Slovak Republic}

\begin{abstract} 
We calculated spectral energy distribution of the helium strong
chemically peculiar star HD 37776 using adequate model atmospheres.
We show that the chemical peculiarity influences the stellar energy
distribution. Consequently, spots of peculiar chemical composition
on the surface of a rotating star may cause detectable light
variations of the star. However, the observed light variations are
not likely caused by an uneven surface distribution of helium, but
may be due to spots of metals (mainly carbon).
\end{abstract}

\section{Introduction}

\newcommand\hvezda{HD\,37776}

Magnetic chemically peculiar (mCP) stars exhibit periodic light
variations, that used to be explained by presence of ``photometric"
spots with uneven energy distribution on the surface of rotating
stars. These photometric spots are likely connected with the
``spectroscopic" spots of a peculiar abundance of various chemical
elements. The departures of energy distribution in spectra of mCP
stars from normal energy distribution are explained as a consequence
of line and continuum blanketing originating in the spectroscopic
spots with peculiar chemical composition \citep[e.g.][c.f.
\citeauthor{towog} \citeyear{towog}]{peter}.

\begin{figure}[t]
\begin{center}
\resizebox{0.90\hsize}{!}{\includegraphics{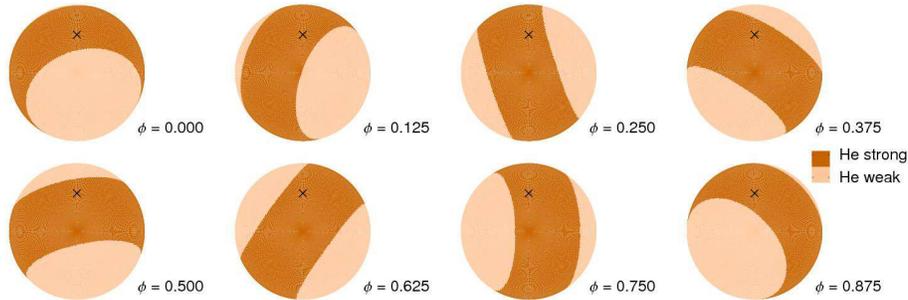}}
\end{center}
\caption{Helium spectroscopic spots as seen at various rotational phases.
Crosses denote the rotational pole. According to
model of \citet{bupo}
there are two helium-poor spots and helium rich ring on the surface of
\hvezda. Phases are calculated according to
\citet{takymy}.}
\label{povrch}
\end{figure}

To test whether the photometric spots are related to the spectroscopic
ones we calculated light variations of the helium strong mCP star \hvezda,
for which \citet{bupo}
by means of Doppler analysis derived a spot model with a helium-rich ring
surrounded by two helium-poor spherical caps (Fig.~\ref{povrch}).

\section{Model atmospheres and calculated flux variations}

\newcommand{\zav}[1]{\left(#1\right)}

For LTE atmosphere modelling we opted code TLUSTY
\citep{hublad}. The models correspond to the helium rich/poor
surface elements on \hvezda\ with
${N(\text{He})/N(\text{H})=0.72}$ and ${0.02}$  respectively
\citep{bupo}. The effective temperature ${{T}_\mathrm{eff}}$ and
the surface gravity ${\log g}$ of all these models are assumed to
be the same and equal to ${{T}_\mathrm{eff}}={22\,000}$\,K and
${{\log g}=4.0}$, respectively \citep{groka}.

The emergent radiative flux was calculated with
SYNSPEC code. The spectral
energy distribution derived for individual surface elements
was used to obtain corresponding fluxes in individual colours.
Finally, these fluxes are integrated
over all visible surface
and the observed magnitude difference is calculated.

\begin{figure}[htb]
\begin{center}
\resizebox{0.61\hsize}{!}{\includegraphics{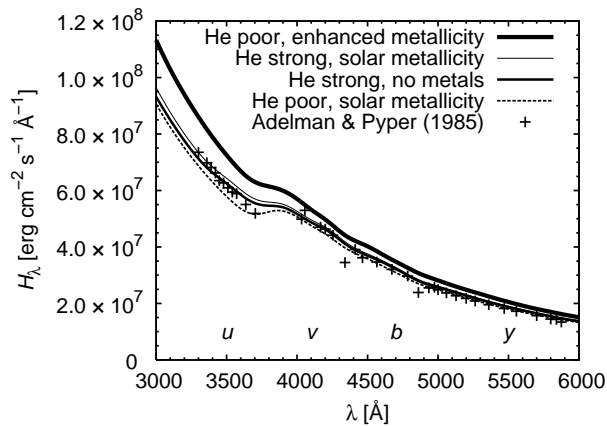}}
\end{center}
\caption{Calculated radiative flux from \hvezda\ atmospheres with different
chemical composition and the observed one \citep{adelpy}}
\label{toky}
\end{figure}

The radiative fluxes calculated for atmospheres with
different chemical composition are displayed on (Fig.~\ref{toky}).
The flux calculated was convolved with a Gauss function to better
demonstrate possible light variations. Apparently, the He-poor surface
elements
with solar metallicity produce a flux that is lower in the visible region
than the flux produced by the He-strong ones. The regions with enhanced
metallicity produces a flux highest in the visible region.

\section{Observed spectrum variations and predicted light variability}

\begin{figure}[ht]
\begin{center}
\resizebox{0.49\hsize}{!}{\includegraphics{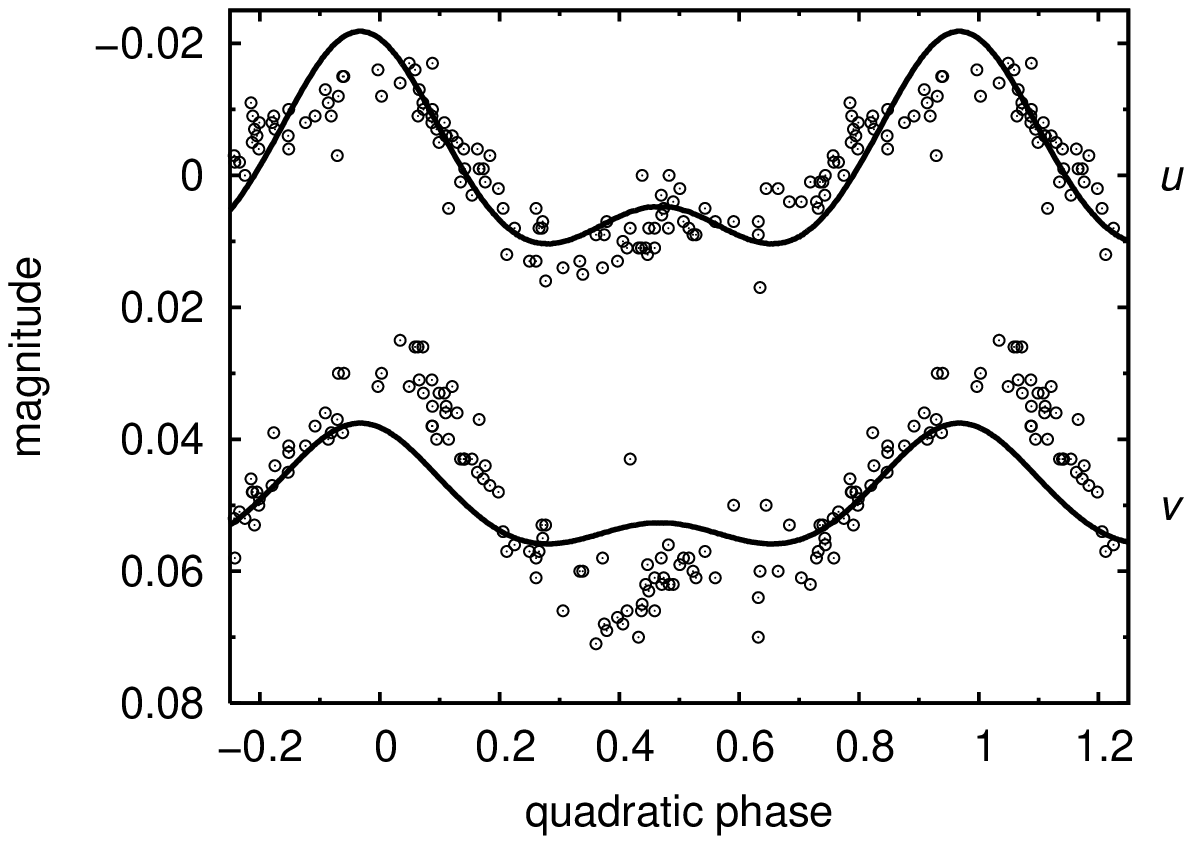}}
\resizebox{0.49\hsize}{!}{\includegraphics{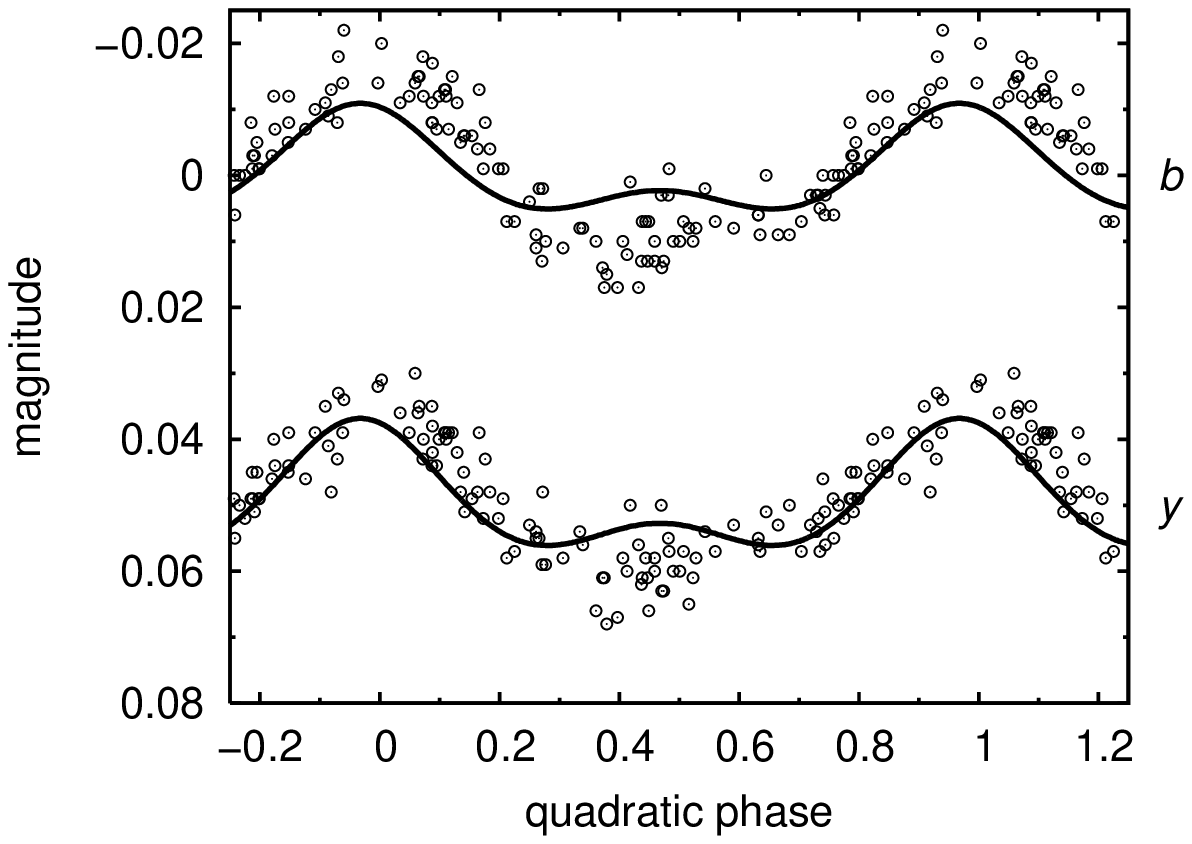}}
\end{center}
\caption{Comparison of predicted (solid line) and observed
\citep[][dots]{takymy} light variability}
\label{hvvel}
\end{figure}

From the spectrum and light variability observations follows that
at the maximum stellar brightness, the helium
equivalent width have its minimum, whereas the Si equivalent
widths have their maximum. Clearly, the He-rich ring is dimmer and
mostly Si-poor, whereas the He-poor spots are brighter and
mostly Si-rich. However, the radiative flux from the solar
metallicity helium-poor surface elements in the visible region  is
lower than that from the helium-strong ones (Fig.~\ref{toky}). Consequently,
the observed light variability of \hvezda\ cannot be explained merely by
altering of emergent flux due to the uneven surface distribution of helium.
Also surface variations of metallicity should be taken in to account.
Therefore, we calculated the predicted light variability of
\hvezda\ assuming that He-poor spots have metallicity thirty
times higher relative to the solar one and that the He-strong ring consists
from H and He only (Fig.~\ref{hvvel}).

\section{Conclusions}

We calculated light curves of the He chemically peculiar star
\hvezda. It is possible to explain its observed
light variability provided that the He-poor spots
have enhanced metallicity whereas the He-strong ring has low
metallicity. In such case the observed light variations are  mostly
due to bound-free transitions of carbon. Although this picture
roughly corresponds to the observed data, detailed spectrum
observations and model improvement are necessary to test our ideas.

\acknowledgements
This work was supported by grants
GA\,\v{C}R 205/03/ D020, 205/04/1267, VEGA 3014 and MVTS SR-\v{C}R
128/04.

\end{document}